\documentclass[]{spie}  

\usepackage{amsmath,amsfonts,amssymb}
\usepackage{subcaption}
\usepackage{graphicx}
\usepackage{cite} 
\usepackage{times}
\usepackage{epsfig}
\usepackage{amsmath}
\usepackage{nccmath}
\usepackage{amssymb}
\usepackage{mwe}
\usepackage{acro}
\usepackage{amssymb}
\usepackage{xcolor,colortbl}
\usepackage{tabularx}
\usepackage{relsize}
\usepackage{pifont}
\usepackage{booktabs} 
\usepackage{multirow}
\usepackage{multicol}
\usepackage{adjustbox}
\usepackage{float}
\usepackage{graphicx}
\usepackage{makecell}
\usepackage{tabu}
\usepackage[colorlinks=true, allcolors=blue]{hyperref}
\usepackage[capitalize]{cleveref}
\usepackage{enumitem}

\title{Evaluation Kidney Layer Segmentation on Whole Slide Imaging using Convolutional Neural Networks and Transformers}

\author[a]{Muhao Liu}
\author[b]{Chenyang Qi}
\author[c]{Shunxing Bao}
\author[a]{Quan Liu}
\author[a]{Ruining Deng}
\author[d]{Yu Wang}
\author[d]{Shilin Zhao}
\author[c]{Haichun Yang}
\author[a,b,e]{Yuankai Huo}

\affil[a]{Department of Computer Science, Vanderbilt University, Nashville, TN, USA}
\affil[b]{Department of Nephrology, Shanghai Tenth People’s Hospital, Tongji University, Shanghai, China}
\affil[c]{Department of Electrical and Computer Engineering, Vanderbilt University, Nashville, TN, USA}
\affil[d]{Department of Biostatistics, Vanderbilt University Medical Center, Nashville, TN, USA}
\affil[e]{Department of Pathology, Microbiology and Immunology, Vanderbilt University Medical Center, Nashville, TN, USA}

\authorinfo{Corresponding author: Yuankai Huo: E-mail: yuankai.huo@vanderbilt.edu}

\begin{document} 
\maketitle

\begin{abstract}
The segmentation of kidney layer structures, including cortex, outer stripe, inner stripe, and inner medulla within human kidney whole slide images (WSI) plays an essential role in automated image analysis in renal pathology. However, the current manual segmentation process proves labor-intensive and infeasible for handling the extensive digital pathology images encountered at a large scale. In response, the realm of digital renal pathology has seen the emergence of deep learning-based methodologies. However, very few, if any, deep learning based approaches have been applied to kidney layer structure segmentation. Addressing this gap, this paper assesses the feasibility of performing deep learning based approaches on kidney layer structure segmetnation. This study employs the representative convolutional neural network (CNN) and Transformer segmentation approaches, including Swin-Unet, Medical-Transformer, TransUNet, U-Net, PSPNet, and DeepLabv3+. We quantitatively evaluated six prevalent deep learning models on renal cortex layer segmentation using mice kidney WSIs. The empirical results stemming from our approach exhibit compelling advancements, as evidenced by a decent Mean Intersection over Union (mIoU) index. The results demonstrate that Transformer models generally outperform CNN-based models. By enabling a quantitative evaluation of renal cortical structures, deep learning approaches are promising to empower these medical professionals to make more informed kidney layer segmentation.
\end{abstract}



  


\section{INTRODUCTION}
\label{sec:intro}  

Digital pathology has proven its prowess in quantifying digitized tissues using whole slide images (WSIs). It is acknowledged for facilitating remote analysis of diseased tissues, replacing labor-intensive manual pathology assessments~\cite{huo2021ai} with computer-assisted methodologies. Region segmentation is of particular significance~\cite{freixenet2002yet}, a specialized branch of semantic segmentation, vital for precise diagnosis and treatment. Yet, the task of segmenting WSIs, particularly the intricate tissue structure of the kidney layer structures, presents challenges due to cell similarity across distinct regions.


This study aims to bridge this gap by conducting a comparative analysis of three widely used CNN models (U-Net~\cite{ronneberger2015u}, PSPNet~\cite{zhao2017pyramid}, DeepLabv3+~\cite{chen2018encoder}) and three Transformer models (Swin-Unet~\cite{cao2022Swin}, TransUNet~\cite{chen2021transunet}, Medical-Transformer~\cite{valanarasu2021medical}) on kidney layer segmentation task. We compare the widely used approach to offer an overview of the functioning of deep learning in kidney layer segmentation. To our knowledge, we are among the pioneers in evaluating deep learning's segmentation capabilities for kidney layer structures.  


To underscore the validation aspect of this research, we conducted a comprehensive series of experiments, each designed with precision and thoroughness, including: comprehensive evaluations of different models, comparing CNNs and Transformer models, testing against varying complexities within the dataset. By comparing the effectiveness of CNNs and Transformer models and applying them to mice kidney cortex segmentation, our study not only opens avenues for future research in this realm but also provides references for upcoming researchers grappling with model selection in kidney layer segmentation applications.

\section{METHOD}
\subsection{Overall framework}
\begin{figure*}[t]
\begin{center}
\includegraphics[width=0.8\linewidth]{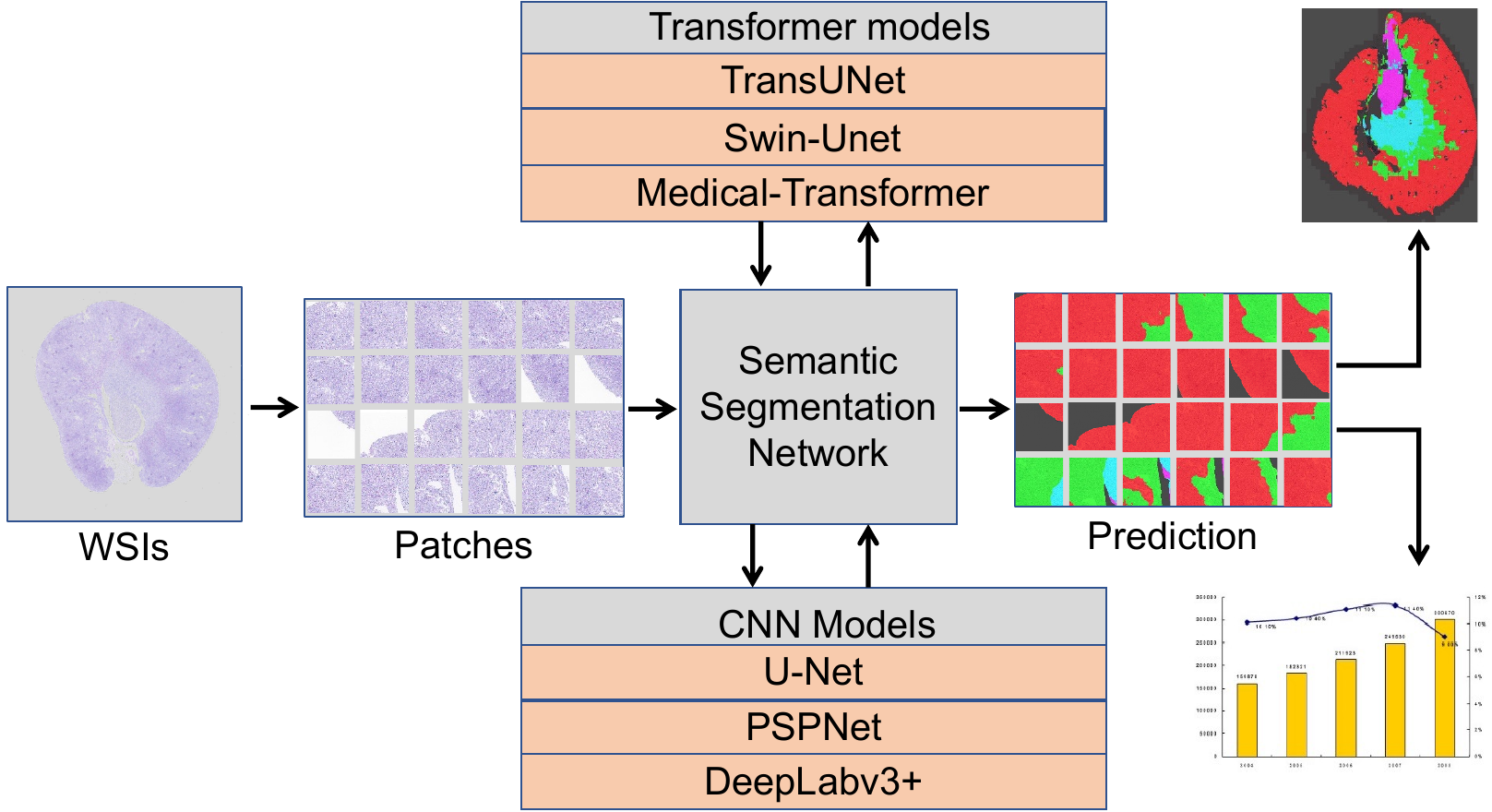}
\end{center}
\caption{This figure shows general steps of the experiment. The WSIs are first tiled into patches. Then, six prevalent deep learning segmentation approaches are applied to all patches. Last, the patch-level results are assembled to WSI-level.}
\label{fig:pipeline}
\end{figure*}

Fig.\ref{fig:pipeline} shows the overall framework of this study, including five main steps: (1) Annotations are imported onto the tissue image, exported as labels in GeoJSON format, and simultaneously, tissue images are exported in JPG format. (2) The exported labels are assigned different pixel values. Both labels and whole slide images (WSIs) are then downsampled by a factor of 3. Subsequently, the view is segmented into patches of 1024$\times$1024 pixels, using a half-step sliding window for smoother results. Patches with no pixel information are discarded, and incomplete patches are resized to maintain uniformity. (3) The mage patches and label patches are matched and collated. (4) The collated patches are fed into various models for training, with temporary best-performing models saved every 10 epochs for potential later use. (5) The top-performing model is then used for prediction on new patches of size 1024$\times$1024, downsampled by a factor of 1. The model's predictions are overlaid and merged, yielding the final predicted WSIs.

\begin{table}[h!]
\centering
\caption{CNN-Based Models}
\begin{tabular}{p{3cm}cccc}
\toprule
Model & Acronym & Year & Backbone & Download URL \\
\midrule
U-Net~\cite{ronneberger2015u} & U-Net & 2015 & VGG~\cite{simonyan2014very} & \href{https://github.com/milesial/Pytorch-UNet}{Link} \\
\midrule
DeepLabv3+~\cite{chen2018encoder} & v3+ & 2018 & MobileNetV2~\cite{sandler2018mobilenetv2} & \href{https://github.com/VainF/DeepLabV3Plus-Pytorch}{Link} \\
\midrule
Pyramid Scene Parsing Net~\cite{zhao2017pyramid} & PSPNet & 2016 & ResNet50~\cite{he2016deep} & \href{https://pypi.org/project/segmentation-models-pytorch/}{Link} \\
\bottomrule
\end{tabular}
\label{table:models}
\end{table}

\begin{table}[h!]
\centering
\caption{Transformer-Based Models}
\begin{tabular}{p{3cm}cccc}
\toprule
Model & Acronym & Year & Backbone & Download URL \\
\midrule
TransUNet~\cite{chen2021transunet} & TUnet & 2021 & Vision Transformer~\cite{dosovitskiy2020image} & \href{https://github.com/Beckschen/TransUNet}{Link} \\
\midrule
Swin-Unet~\cite{cao2022Swin} & SUnet & 2021 & Swin Transformer~\cite{liu2021Swin} & \href{https://github.com/HuCaoFighting/Swin-Unet}{Link} \\
\midrule
Medical-Transformer~\cite{valanarasu2021medical} & MedT & 2021 & Gated Axial-Attention~\cite{valanarasu2021medical} & \href{https://github.com/jeya-maria-jose/Medical-Transformer}{Link} \\
\bottomrule
\end{tabular}
\label{table:transformer_models}
\end{table}

\subsection{CNN-Based methods}
Three CNNs based models and three Transformer based models, which open-source code is available, are utilized in this study:
\begin{enumerate}
\item \textbf{U-Net~\cite{ronneberger2015u}}:  This network has a U-shaped design that uses convolutions, pooling, and up-convolutions. While it can handle segmentation tasks and learn from few labeled images, it requires a lot of computing power and can be expensive.
\item \textbf{PSPNet~\cite{zhao2017pyramid}}:  This network uses a pyramid pooling module for segmentation, which may enhance performance under complex conditions. It excels at grasping the overall context without overlooking important details as it processes various regions of the image. However, it demands significant computational resources, which can increase memory usage and slow down processing.
\item \textbf{DeepLabv3+~\cite{chen2018encoder}}:  It employs atrous convolutions and atrous spatial pyramid pooling (ASPP) to grasp the image context across various scales. Atrous convolution enables effectively capturing a broader context without adding computational demands. While ASPP integrates different scales, this increases the computational load, balancing out the benefits of using atrous convolution.
\end{enumerate}

\subsection{Transformer-Based methods}
\begin{enumerate}
\item \textbf{TransUNet~\cite{chen2021transunet}}: This framework uses a mix of CNN-Transformer architecture, blending the global contextualization benefits of transformers with the high-resolution spatial information strengths of CNNs.

\item \textbf{Swin-UNet~\cite{cao2022Swin}}: This is a U-shaped architecture solely based on Transformers, comprising an encoder, bottleneck, decoder, and skip connection. Images are divided into distinct, non-overlapping patches and channeled into the Transformer encoder. Afterward, the contextual features learned are up-sampled.

\item \textbf{Medical-Transformer~\cite{valanarasu2021medical}}: This network utilizes the gated position-sensitive axial attention mechanism and the Local-Global training approach. While it processes the whole image, it also zeroes in on intricate details. It's recognized for its effectiveness and strong performance, even with smaller datasets.
\end{enumerate}

\section{DATA AND EXPERIMENT}

\subsection{Data}
The datasets consist of a training set and a validation set, as well as an external testing set. The training set comprises 8 wild-type mice, 10 db/db mice on the BKS background, which serve as models for type II diabetes, and 12 db/db mice treated with an angiotensin receptor blocker (ARB). These mice were exclusively male and aged between 20 to 24 weeks. Conversely, the validation set comprises young adult female mice aged 11 to 14 weeks, along with aging female mice aged between 21 to 24 months.
Upon harvesting, the kidneys were fixed in paraformaldehyde, and subsequently, 3\textmu m sections were cut and subjected to staining using Periodic acid-Schiff (PAS). Whole-slide images (WSIs) were captured under a 40$\times$ magnification. We split the entire training dataset containing 30 WSIs with a ratio of 7:3 for training and validation. We withheld 25 WSIs as testing set. Subsequently, both labels and WSIs were cut using a size of 1024$\times$1024, using a half-step sliding window approach (half of the patch size in both horizontal and vertical directions) to ensure smoother results. If a complete patch cannot be extracted due to boundary restraints, the resulting patch was resized to the size of 1024$\times$1024. To improve data quality, the post-condition was set to exclude all patches containing no pixel value. 

\subsection{Experiment environment and procedures}
Annotations were imported into QuPath, superimposed onto issue images, and subsequently exported as labels in GeoJSON format. Concurrently, tissue images were exported into JPG format. Then, exported labels were then filled with distinct pixel values, and both labels and WSIs were downsampled by a factor of 3$\times$. The colors of masks and pixel values for each region were presented in Table. Each region has two pixel values assigned for training and visualization.

Following data preparation, image patches and their associated label patches were collated following the structure of image standard segmentation practice~\cite{everingham2010pascal}, a widely recognized benchmark for image classification and object detection tasks. During training, both image and label patches were imported and downsampled online. Each model underwent training over 100 epochs, during which the entire set of approximately 4000 training patches were fed into the network in batches of 8. After each epoch, the total loss (Total\_Loss) and validation loss (Val\_Loss) were calculated, with the model achieving the lowest validation loss being saved for later evaluation. The performance of this model was then assessed against the complete test dataset using the mean Intersection over Union (mIoU) metric, as defined in Equation \ref{eq:mIoU}. Alongside mIoU, it is equally imperative to consider the Dice Similarity Coefficient (DSC), a critical metric to evaluate the spatial overlap accuracy between the predicted segmentation and the manual segmentation, ensuring a comprehensive evaluation of the model's performance. This score would further provide a robust assessment of the segmentation accuracy. Furthermore, to corroborate the robustness of each model, tests were conducted on a unseen dataset of 20 whole slide images (WSIs). The DSC indices generated from these WSIs for each model were employed to calculate descriptive statistical measures, including mean, median, and standard deviation, shown in the Table~\ref{table:model_performances}. These measures furnished a nuanced understanding of the performance consistency and reliability exhibited by each model. All relevant data have been tabulated and will be discussed in the following sections.


\begin{equation}
\text{mIoU} = \frac{1}{n} \sum \frac{A \cap B}{A \cup B}
\label{eq:mIoU}
\end{equation}

\noindent where $n$ is the number of classes. $A$ is the area of prediction for a given class. $B$ is the area of the manual segmentation for that class.
$\cap$ represents intersection (common area between prediction and manual segmentation). $\cup$ represents union (total area covered by both prediction and manual segmentation). $\sum$ represents the sum over all classes~\cite{everingham2010pascal}. mIoU index is often used as a standard of performance by models in semantic segmentation, given its ability to take both false positives and false negatives into account, providing a comprehensive view of the model’s performance. We compared the above three CNNs with these three transformer models.

\begin{table}[h!]
\centering
\caption{Comparison of Model Performances (\%)}
\begin{tabular}{p{4cm}cccc}
\toprule
Model & mIoU & median DSC & mean DSC & standard deviation DSC \\
\midrule
U-Net & 83.7 & 65.5 & 61.8 & 14.2 \\
\midrule
PSPNet & 83.7 & 71.0 & 67.5 & 15.9 \\
\midrule
DeepLabv3+ & 81.6 & 61.9 & 60.9 & 14.9 \\
\midrule
MedT & 88.5 & 72.3 & 70.1 & 17.0 \\
\midrule
Swin-Unet & \textbf{92.2} & \textbf{81.0} & \textbf{77.4} & 14.1 \\
\midrule
Trans-Unet & 91.9 & 80.1 & 76.0 & \textbf{12.8} \\
\bottomrule
\end{tabular}
\label{table:model_performances}
\end{table}

\begin{figure}[h]
\centering
\includegraphics[width=0.8\textwidth]{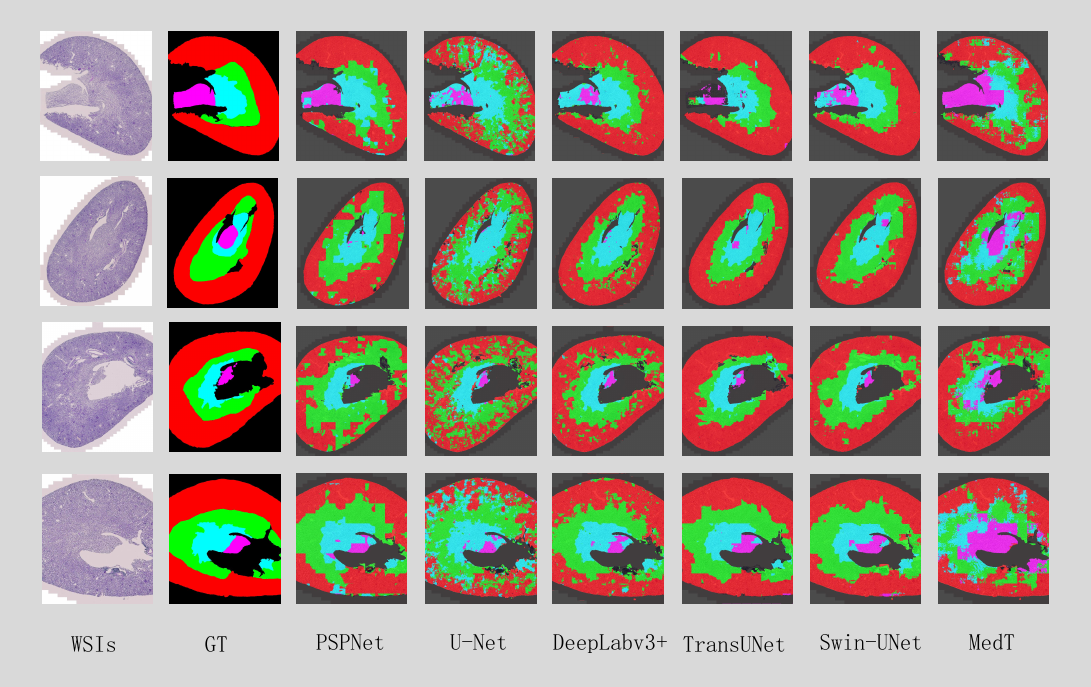}
\caption{This figure shows comparisons between manual segmentation and predictions.}
\label{fig:gt_prediction}
\end{figure}

\begin{figure}[h]
\begin{center}
\includegraphics[width=0.7\linewidth]{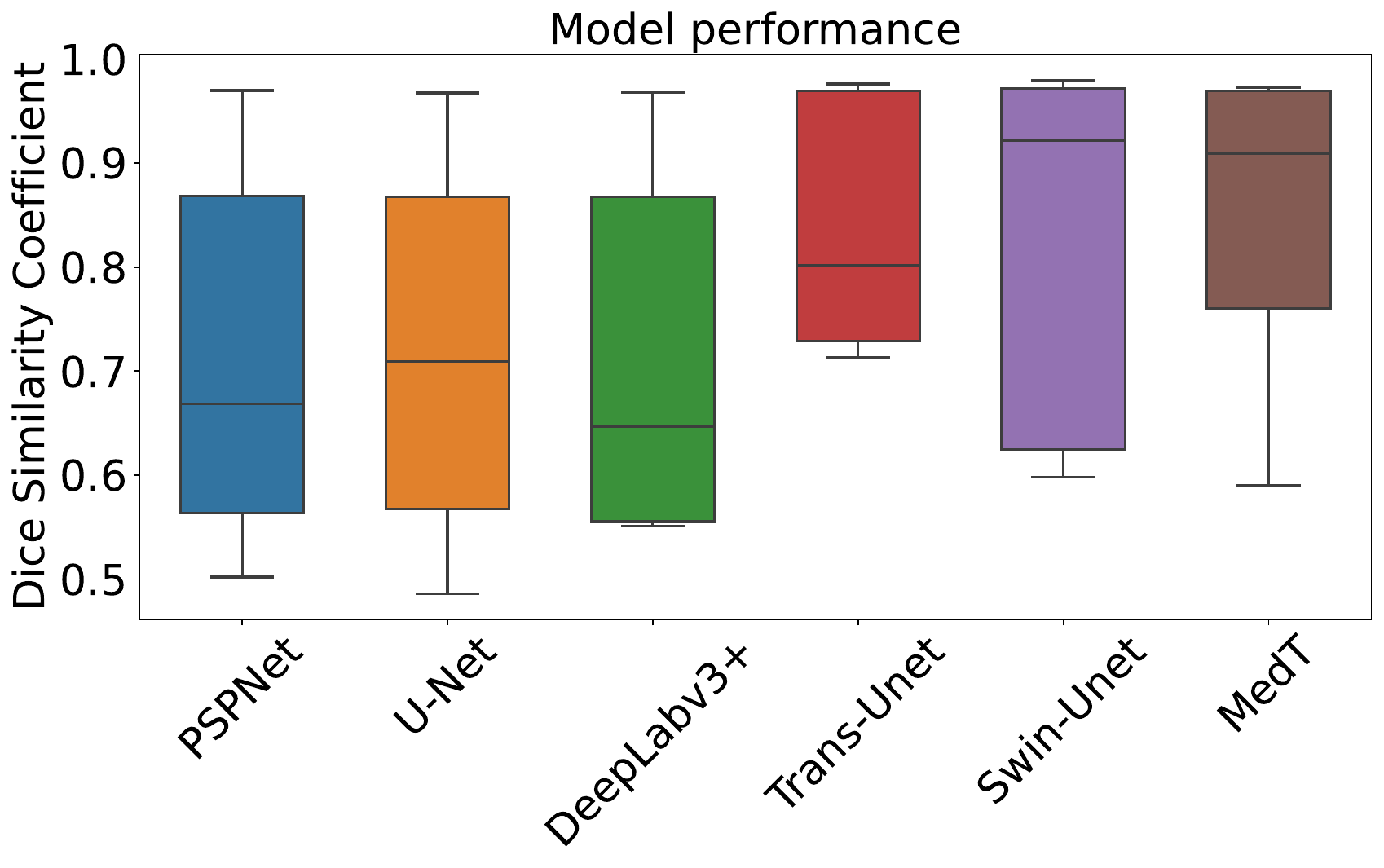}
\end{center}
\caption{This figure shows mIoU scores on 20 WSIs of all six models}
\label{fig:boxplot}
\end{figure}


\section{RESULTS}
The performance metrics of each of the selected models for kidney layer segmentation are presented in Table~\ref{table:models} and Table~\ref{table:transformer_models}. 
Each mIoU index displayed in Table~\ref{table:model_performances} reported represents an average across all ten test slides. The DSC of six models is visualized in Fig.~\ref{fig:boxplot}. Additionally, Fig.~\ref{fig:gt_prediction} provides a comparison of the predicted WSIs against their corresponding manual segmentation for four selected test slides. Based on mIoU index values, three Transformer models in this study generally outperform three CNN models. Among transformer models, Swin-Unet exhibits the best performance with an accuracy of 92.2\%. This superior performance may be attributed to the model's unique architecture which effectively combines the benefits of the Transformer's global attention mechanism with the strong local feature extraction ability of U-Net. Moreover, Trans-Unet has the most stable performance as indicated by the standard deviation of 12.8. During the training process, Trans-Unet could have reached a better, more stable region in the solution space, leading to more consistent results. It's also possible that the specific features of the kidney mice cortex dataset are particularly well-suited to the Trans-Unet model, leading to more consistent performance. Furthermore, while Swin-Unet is purely Transformer-based network, Trans-Unet employs a hybrid CNN-Transformer architecture. It means that whether hybrid or purely Transformer-based architectures is not a single influential factor in the model's performance.

\section{Conclusions}
This research conducts a comparison between prevalent CNN and Transformer-based segmentation algorithms for kidney layer segmentation. The outcomes reveal that Transformer Models generally outperform CNNs in terms of performance. Notably, the Swin-Unet model attains the highest overall accuracy. This investigation underscores the significance of selecting appropriate models when handling complex datasets, showcasing the promising capabilities of Transformer models. It is anticipated that this study will serve as a point of reference for forthcoming research on the novel task of kidney layer segmentation, thus providing valuable assistance to future endeavors in this domain.

\section{ACKNOWLEDGMENTS}       
This work has not been submitted for publication or presentation elsewhere. This work is supported in part by NIH R01DK135597(Huo), DoD HT9425-23-1-0003(HCY), and NIH NIDDK DK56942(ABF).

\bibliography{main} 

\begin{thebibliography}{10}

\bibitem{huo2021ai}
Huo, Y., Deng, R., Liu, Q., Fogo, A.~B., and Yang, H., ``Ai applications in
  renal pathology,'' {\em Kidney international}~{\bf 99}(6),  1309--1320
  (2021).

\bibitem{freixenet2002yet}
Freixenet, J., Munoz, X., Raba, D., Mart{\'\i}, J., and Cuf{\'\i}, X., ``Yet
  another survey on image segmentation: Region and boundary information
  integration,'' in [{\em Computer Vision—ECCV 2002: 7th European Conference
  on Computer Vision Copenhagen, Denmark, May 28--31, 2002 Proceedings, Part
  III 7}{\nolinebreak\hspace{0.1em}]},   408--422, Springer (2002).

\bibitem{ronneberger2015u}
Ronneberger, O., Fischer, P., and Brox, T., ``U-net: Convolutional networks for
  biomedical image segmentation,'' in [{\em Medical Image Computing and
  Computer-Assisted Intervention--MICCAI 2015: 18th International Conference,
  Munich, Germany, October 5-9, 2015, Proceedings, Part III
  18}{\nolinebreak\hspace{0.1em}]},   234--241, Springer (2015).

\bibitem{zhao2017pyramid}
Zhao, H., Shi, J., Qi, X., Wang, X., and Jia, J., ``Pyramid scene parsing
  network,'' in [{\em Proceedings of the IEEE conference on computer vision and
  pattern recognition}{\nolinebreak\hspace{0.1em}]},   2881--2890 (2017).

\bibitem{chen2018encoder}
Chen, L.-C., Zhu, Y., Papandreou, G., Schroff, F., and Adam, H.,
  ``Encoder-decoder with atrous separable convolution for semantic image
  segmentation,'' in [{\em Proceedings of the European conference on computer
  vision (ECCV)}{\nolinebreak\hspace{0.1em}]},   801--818 (2018).

\bibitem{cao2022Swin}
Cao, H., Wang, Y., Chen, J., Jiang, D., Zhang, X., Tian, Q., and Wang, M.,
  ``Swin-unet: Unet-like pure transformer for medical image segmentation,'' in
  [{\em European conference on computer vision}{\nolinebreak\hspace{0.1em}]},
  205--218, Springer (2022).

\bibitem{chen2021transunet}
Chen, J., Lu, Y., Yu, Q., Luo, X., Adeli, E., Wang, Y., Lu, L., Yuille, A.~L.,
  and Zhou, Y., ``Transunet: Transformers make strong encoders for medical
  image segmentation,'' {\em arXiv preprint arXiv:2102.04306}  (2021).

\bibitem{valanarasu2021medical}
Valanarasu, J. M.~J., Oza, P., Hacihaliloglu, I., and Patel, V.~M., ``Medical
  transformer: Gated axial-attention for medical image segmentation,'' in [{\em
  Medical Image Computing and Computer Assisted Intervention--MICCAI 2021: 24th
  International Conference, Strasbourg, France, September 27--October 1, 2021,
  Proceedings, Part I 24}{\nolinebreak\hspace{0.1em}]},   36--46, Springer
  (2021).

\bibitem{simonyan2014very}
Simonyan, K. and Zisserman, A., ``Very deep convolutional networks for
  large-scale image recognition,'' {\em arXiv preprint arXiv:1409.1556}
  (2014).

\bibitem{sandler2018mobilenetv2}
Sandler, M., Howard, A., Zhu, M., Zhmoginov, A., and Chen, L.-C.,
  ``Mobilenetv2: Inverted residuals and linear bottlenecks,'' in [{\em
  Proceedings of the IEEE conference on computer vision and pattern
  recognition}{\nolinebreak\hspace{0.1em}]},   4510--4520 (2018).

\bibitem{he2016deep}
He, K., Zhang, X., Ren, S., and Sun, J., ``Deep residual learning for image
  recognition,'' in [{\em Proceedings of the IEEE conference on computer vision
  and pattern recognition}{\nolinebreak\hspace{0.1em}]},   770--778 (2016).

\bibitem{dosovitskiy2020image}
Dosovitskiy, A., Beyer, L., Kolesnikov, A., Weissenborn, D., Zhai, X.,
  Unterthiner, T., Dehghani, M., Minderer, M., Heigold, G., Gelly, S., et~al.,
  ``An image is worth 16x16 words: Transformers for image recognition at
  scale,'' {\em arXiv preprint arXiv:2010.11929}  (2020).

\bibitem{liu2021Swin}
Liu, Z., Lin, Y., Cao, Y., Hu, H., Wei, Y., Zhang, Z., Lin, S., and Guo, B.,
  ``Swin transformer: Hierarchical vision transformer using shifted windows,''
  in [{\em Proceedings of the IEEE/CVF international conference on computer
  vision}{\nolinebreak\hspace{0.1em}]},   10012--10022 (2021).

\bibitem{everingham2010pascal}
Everingham, M., Van~Gool, L., Williams, C.~K., Winn, J., and Zisserman, A.,
  ``The pascal visual object classes (voc) challenge,'' {\em International
  journal of computer vision}~{\bf 88},  303--338 (2010).

\end{thebibliography}
\bibliographystyle{spiebib} 

\end{document}